\documentclass[conference,a4paper]{IEEEtran}
\IEEEoverridecommandlockouts
\usepackage{theorem}
\usepackage{cite}
\usepackage{amsmath,amssymb,amsfonts}
\usepackage{algorithmic}
\usepackage{graphicx}
\usepackage{textcomp}
\usepackage{xcolor}
\def\BibTeX{{\rm B\kern-.05em{\sc i\kern-.025em b}\kern-.08em
    T\kern-.1667em\lower.7ex\hbox{E}\kern-.125emX}}

{\theorembodyfont{\rmfamily}
   \newtheorem{The}{{\textbf Theorem}}}
    {\theorembodyfont{\rmfamily}
   \newtheorem{Lem}{{\textbf Lemma}}}
     {\theorembodyfont{\rmfamily}
   \newtheorem{Cor}{{\textbf Corollary}}}
    
\begin{document}

%*********************AEIT COPYRIGHT****************************
\makeatletter
\def\ps@IEEEtitlepagestyle{%
  \def\@oddfoot{\mycopyrightnotice}%
  \def\@evenfoot{}%
}
\def\mycopyrightnotice{%
  {\footnotesize 978-88-87237-53-5 \copyright 2022 AEIT \hfill} 
}
\makeatother
%***************************************************************

\title{Punctured Binary Simplex Codes as LDPC codes}

\author{\IEEEauthorblockN{Massimo Battaglioni}
\IEEEauthorblockA{\textit{Dept. of Information Engineering} \\
\textit{Marche Polytechnic University}\\
Ancona, Italy \\
m.battaglioni@staff.univpm.it}
\and
\IEEEauthorblockN{Giovanni Cancellieri}
\IEEEauthorblockA{\textit{Dept. of Information Engineering} \\
\textit{Marche Polytechnic University}\\
Ancona, Italy \\
g.cancellieri@staff.univpm.it}}

\maketitle

\begin{abstract}
Digital data transfer can be protected by means of suitable error correcting codes. Among the families of state-of-the-art codes, LDPC (Low Density Parity-Check) codes have received a great deal of attention recently, because of their performance and flexibility of operation, in wireless and mobile radio channels, as well as in cable transmission systems. In this paper, we present a class of rate-adaptive LDPC codes, obtained as properly punctured simplex codes. These codes allow for the use of an efficient soft-decision decoding algorithm, provided that a condition called row-column constraint  is satisfied. This condition is tested on small-length codes, and then extended to medium-length codes. The puncturing operations we apply do not influence the satisfaction of the row-column constraint, assuring that a wide range of code rates can be obtained. We can reach code rates remarkably higher than those obtainable by the original simplex code, and the price in terms of minimum distance turns out to be relatively small, leading to interesting trade-offs in the resulting asymptotic coding gain. 
\end{abstract}

%Block lengths of the order of 500, at code rates from 1/4 to 1/2, appear suitable for practical applications.


\section{Introduction}
\begin{figure}[htbp]
\centering
    \includegraphics[width=0.35\textwidth]{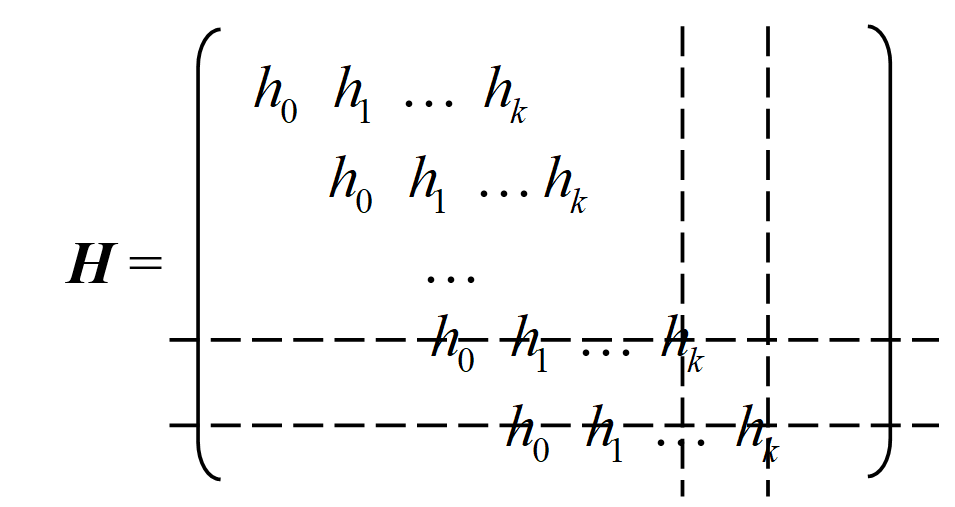}
\caption{General form of the parity-check matrix of (punctured) simplex codes. Out of the main diagonal band, only $0$ symbols are present. 
Effects of $s = 2$ row-column eliminations, starting from the bottom right.}
\label{fig:Hgeneral}
\end{figure}

Simplex codes are duals of Hamming codes \cite{sloane}. In polynomial representation, for a binary finite field, they exhibit a parity-check matrix $\mathbf{H}$ where a primitive binary polynomial $h(x)$ shifts along a diagonal trace, from left to right, by one position each row. On the cyclic code length $N$, the tern describing simplex codes is $[N,k,d]$, where $N=2^k-1$ is the block length and $d_{\min}=2^{k-1}$ is the code minimum distance. All the $2^{k}-1$ non-null codewords have weight $d_{\min}$ \cite{Peterson}. Precisely, such non-null code words represent all the possible cyclic shifts of the same maximum-length pseudo-random binary sequence \cite{Golomb}.
The parity-check polynomial $h(x)$ has degree $k$, with coefficients $h_0=h_k=1$, and can be interpreted as the generator polynomial of a Hamming code (the dual of our code) having the same cyclic code length $N$. In Fig. \ref{fig:Hgeneral} the general form of such an $\mathbf{H}$ matrix is shown. It exhibits $N$ columns and $N-k$ rows.

It is possible to choose a shorter code length $n$, with $n < N$, by eliminating external rows and hence external columns. This operation is called puncturing \cite{Costello} and can be repeated as many times as one wishes. So $n$ becomes a variable, whereas $k$ does not change. After $s$ elementary row-column eliminations, the punctured code is described by the new tern $[n,k,d']$, with $n = N-s$, and the number of rows in the new parity-check matrix $\mathbf{H}'$ is reduced to $r = n-k$. Finally, $d'$ is the new minimum distance. After this construction, the rows of the parity-check matrix remain all linearly independent, so that $\mathbf{H}'$  still has full rank.

The code rate $\frac{k}{n}$ can be easily synthesized, leading to a rate-adaptive coding system. It is unavoidable that the new minimum distance $d'$ becomes smaller and smaller, for increasing values of $s$. Nevertheless, the main drawback of simplex codes, on their cyclic length $N$, that is a very low code rate, can be partially overcome.
The problem of predicting word weight distributions and new minimum distances $d’$ for punctured binary simplex codes has been already faced \cite{BaldiSimplex,Shirvanimoghaddam}. Nevertheless, a research question remains open, regarding an efficient low-complexity soft-decision decoding procedure, able to exploit the good design characteristics of these codes. The main contribution of the present paper is in the interpretation of the parity-check matrix of simplex codes as a sparse matrix. Owing to this, the decoding algorithms which are suitable for Low-Density Parity-Check (LDPC) codes, can be adopted. In this context, the conditions for satisfying row-column constraint \cite{RyanBook} will be investigated, in order to assure a straightforward decoding procedure, e.g. by means of  the sum-product algorithm \cite{Kschischang}. The availability of primitive polynomials to be chosen as $h(x)$ will be verified. Furthermore, a circulant expansion procedure \cite{RyanBook} in designing the final form of the $\mathbf{H}$ matrix will be suggested. Some simulations of the code performance on an Additive White Gaussian Noise (AWGN) channel will demonstrate feasibility of the proposed solution.
The arguments are organized as follows. In Section \ref{sec:preli} we provide some preliminary considerations. In Section \ref{sec:wgen} we obtain some theoretical results. In Section \ref{sec:weidi} some word weight distributions are calculated, allowing to predict the progressive performance improvement for increasing values of $k$. In Section \ref{sec:num} we provide some numerical results, in terms of BER curves. Finally, we draw some concluding remarks in Section \ref{sec:concl}.

\section{Preliminaries}\label{sec:preli}

A Golomb ruler is a sequence of non-negative integers such that every difference of two integers in the sequence is distinct.

The Hamming weight of a vector is defined as the number of non-zero symbols it contains and is simply called weight in the following.

In this paper, we only consider binary LDPC codes. 
LDPC codes are a family of linear codes characterized by parity-check matrices having a relatively small number of non-zero entries compared to the number of zeros. 
Namely, if an LDPC $\mathbf H\in\mathbb F_2^{r\times n}$ has full rank $r<n$ and row and column weight in the order of $\log(n)$ and $\log(r)$, respectively, then it defines an LDPC code with length $n$ and dimension $k = n-r$, with code rate $R=\frac{k}{n}$. If all the rows of $\mathbf{H}$ have the same weight, we denote it as $w_c$. 
The associated code is $C = \left\{\mathbf c\in\mathbb F_q^n | \mathbf c \mathbf H^\top = \mathbf 0 \right\}$, where $^\top$ denotes transposition. The number of codewords of weight $w$ is denoted as $A(w)$.

The row-column constraint in the parity-check matrix of an LDPC code expresses the condition of not having four $1$-symbols in the vertices of a rectangular geometry, forming a $4$-length closed cycle in that matrix. It is well known that  soft-decision decoding algorithms, like the sum-product algorithm, exhibit convergence problems when working on parity-check matrices containing the aforementioned $4$-length cycles.

In the following, we consider punctured simplex codes as LDPC codes, represented by parity-check matrices as those in Fig. \ref{fig:Hgeneral}, described by a primitive parity-check polynomial $h(x)=1+h_1x+\ldots+h_2x^{k-1}+x^k$ of degree $k$ and weight $w$, where $h_i$ is either $0$ or $1$. We define the vector $\mathbf{h}$ containing the $h_i$s, for $i\in\{0,\ldots, k\}$. We also define the vector $\mathbf{p}$ of length $w$, containing $\{i \in \{0,\ldots,k\} | h_i=1 \}$ in ascending order. In other words, $\mathbf{p}$ is the support of the vector containing the coefficients of the polynomial. Finally, we define the vector $\mathbf{s}$ of length $w-1$, such that $s_i=p_{i+1}-p_{i}$, $i\in \{0,\ldots,w-2\}$. The following result holds.

\begin{The}
A necessary and sufficient condition for the satisfaction of the row-column constraint for a punctured simplex code is that the corresponding $\mathbf p$, derived from the primitive parity-check polynomial $h(x)$, is a Golomb ruler.
\label{the:the1}
\end{The}
\begin{IEEEproof}
A $4$-length cycle exists in $\mathbf{H}$ if and only if there exist two pairs $(i_1,i_2)$, $(i_3,i_4)$ such that $h_{i_1}=h_{i_2}=h_{i_3}=h_{i_4}=1$ and $i_2-i_1=i_4-i_3$, being $(i_1,i_2,i_3,i_4)$ different one another, except that it might be $i_1=i_4$. 

Each entry of $\mathbf{p}$ corresponds to a non-zero coefficient of $h(x)$. Then, if $\mathbf{p}$ is a Golomb ruler, by definition, there cannot exist two pairs of different indices $(j_1,j_2)$ and $(j_3,j_4)$ such that $p_{j_2}-p_{j_1}=p_{j_4}-p_{j_3}$. However, if $\mathbf{p}$ contains $p_k$, for some $k$, then, by definition, $h_{p_k}=1$. Therefore, if $\mathbf{p}$ is a Golomb ruler, there cannot exist two pairs $(i_1,i_2)$, $(i_3,i_4)$ such that $h_{i_1}=h_{i_2}=h_{i_3}=h_{i_4}=1$ and $i_2-i_1=i_4-i_3$. This implies that, if $\mathbf{p}$ is a Golomb ruler, $\mathbf{H}$ cannot contain $4$-length cycles and therefore satisfies the row-column constraint.

In order to prove that this condition is necessary we need to show that, if $\mathbf{p}$ is not a Golomb ruler, then $\mathbf{H}$ does not satisfy the row-column constraint. If $\mathbf{p}$ is not a Golomb ruler, then there exist two pairs $(j_1,j_2)$ and $(j_3,j_4)$ such that $p_{j_2}-p_{j_1}=p_{j_4}-p_{j_3}$, also implying that $h_{p_{j_1}}=h_{p_{j_2}}=h_{p_{j_3}}=h_{p_{j_4}}=1$. This is the condition of existence of a $4$-length cycle, which corresponds to the dissatisfaction of the row-column constraint.
\end{IEEEproof}

Now we will consider the properties emerging from an inspection of all the binary primitive polynomials for $k \in \{3, \ldots, 8 \}$. Since they are formed by couples of reciprocal asymmetric polynomials \cite{Peterson}, in Table \ref{tab:Ms} we report only one element for each couple. The notation adopted consists of representing the binary expressions of any polynomial. The number of different polynomials, on average, grows with $k$, but in this very small sample it is possible to recognize various typical well-known properties.

\begin{table*}[h]
\renewcommand{\arraystretch}{1.5}
\caption{All the primitive polynomials (representing also their reciprocals) for $k\in \{3,\ldots,8\}$.}
%$\frac{g_{\mathrm{free}}-2}{2}m_s+1$
\label{table:TabM}
\centering
\begin{tabular}{|c|c|c|}
\hline
$k$ & $N$ & All the primitive polynomials, in binary representation  \\ \hline
$3$ & $7$ & $1101$ \\\hline
$4$ & $15$ & $11001$ \\\hline
$5$ & $31$ & $101001$,	$110111$,	$101111$\\ \hline
$6$ & $63$ & $1100001$,	$1010111$,	$1110011$,	$1101101$\\ \hline
$7$ & $127$ & $11000001$,	$10010001$,	$11110001$,	$10111001$,
				$11100101$,	$11010101$,	$10110101$,	$11111101$,
				$11110111$\\ \hline
				$8$ & $255$ & $100011101$,	$100101011$,	$101100011$,	$101101001$,
				$101100011$,	$111110101$,	$111001111$
 \\\hline
\end{tabular}
\label{tab:Ms}
\end{table*}

The weight $w$ of primitive polynomials is always an odd integer number, not smaller then $3$. For many values of $k$, primitive polynomials with weight $w = 3$ are present. This is not true in few cases, say for $k = 8$, where the minimum weight is $5$. Nevertheless $5$-weight primitive polynomials, as what is known for $k$ up to $10,000$, are always present when $3$-weight polynomials are not \cite{Seroussi98tableof}.
We are interested in fixing conditions able to assure that the row-column constraint is satisfied.

\section{Analysis of the properties of punctured simplex codes}\label{sec:wgen}

In this section we study the properties of the considered codes, first focusing on parity-check polynomials with weight $3$, and then generalizing the obtained results.

\subsection{Codes characterized by parity-check polynomials of weight $3$}\label{sec:w3}

Although the case of a $3$-weight primitive polynomial gives only poor performance, we will investigate this case in detail, with the purpose of understanding the mechanisms of possible low-weight code word existence. The following property holds.

\begin{The}
Given a primitive polynomial of weight $w = 3$, the row-column constraint is always satisfied on a punctured simplex code.
\label{the:the2}
\end{The}
\begin{IEEEproof}
The proof easily follows from the fact that primitive polynomials are always asymmetrical and are characterized by $h_0=h_k=1$. Given this, we have
$$
\mathbf{p}=[0,s_0,k],
$$
where $s_0\neq \frac{k}{2}$ because of the asymmetry. Then, $\mathbf{p}$ is a Golomb ruler characterized by differences $s_0$, $s_1=k-s_0$ and $k$, and the parity-check matrix constructed with $h(x)$ satisfies the row-column constraint, because of Theorem \ref{the:the1}.
\end{IEEEproof}

We can observe how the only degree of freedom in the design of a $k$-degree $3$-weight polynomial is in the choice of the central non-null power coefficient. Many properties characterizing punctured simplex codes of rate $\frac{1}{2}$ are independent of this choice.

\begin{figure}[htbp]
\centering
    \includegraphics[width=0.4\textwidth]{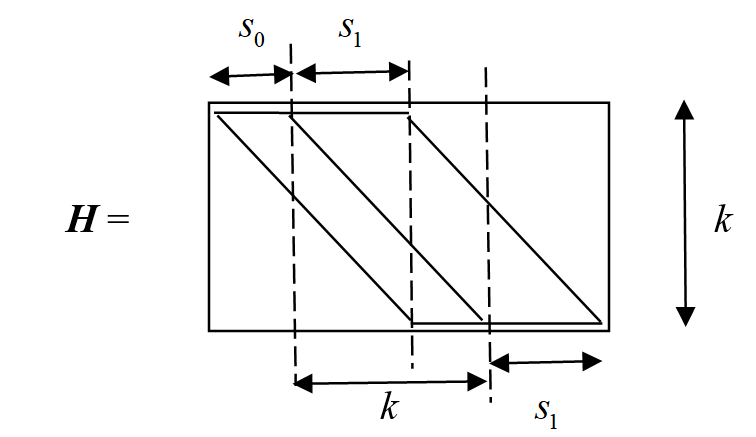}
\caption{Structure of the parity-check matrix for code rate $\frac{1}{2}$ and a $3$-weight parity check polynomial}
\label{fig:H3}
\end{figure}

\begin{Lem}
In a $\frac{1}{2}$-rate punctured simplex code characterized by a $3$-weight parity-check polynomial, among the $2k$ columns of the resulting $\mathbf{H}$ matrix, $k$ have weight $1$ and $k$ have  weight $2$.
\label{lem:lemma12}
\end{Lem}
\begin{IEEEproof}
Using the symbolism in the proof of Theorem \ref{the:the2}, we have the situation depicted in Fig. \ref{fig:H3}, where the diagonal solid lines in the matrix represent symbols 1. Then, we note that the leftmost $s_0$ columns have weight $1$ (and support $0,1,\ldots,s_0-1$, respectively) the central $k$ columns have weight $2$, and the rightmost $s_1$ columns have weight $1$ and support ($s_0=k-s_1,\ldots,k-1$, respectively). Being $s_0+s_1=k$, we have proved the thesis. 

\end{IEEEproof}

Notice that the supports of the $1$-weight columns cover the whole set $\{0,1,\ldots,k-1\}$ without repetitions. This consideration leads to the following result.

\begin{The}
Any $\frac{1}{2}$-rate punctured simplex code, constructed from a $3$-weight parity-check polynomial exhibits minimum distance $d_{\min} = 3$.
\label{the:themindist}
\end{The}
\begin{IEEEproof}
We have proven in Theorem \ref{the:the2} that   $\frac{1}{2}$-rate punctured simplex codes constructed from a $3$-weight parity-check polynomial respect the row-column constraint. In other words, there are no 4-length cycles in the parity-check matrix. Moreover, by construction, all the columns of weight $1$ have a different support. This implies that there cannot exist a pair of columns summing up (modulo $2$) to $0$. Therefore, $d_{\min}>2$. Moreover, given any $2$-weight column in the central portion of $\mathbf{H}$, with support $\{j_1, j_2\}$, by construction (see proof of Lemma \ref{lem:lemma12}) there exist a $1$-weight column on the leftmost portion of $\mathbf{H}$ with support $j_1$ and  a $1$-weight column on the rightmost portion of $\mathbf{H}$ with support $j_2$.  Then, these three columns sum up to zero modulo $2$, implying that $d_{\min}\leq 3$. So, we have $d_{min}=3$. 
\end{IEEEproof}

\begin{Cor}
In the word weight distribution $A(w)$ of a $\frac{1}{2}$-rate punctured simplex code, constructed from a $3$-weight parity-check polynomial, $A(3)=k$ for $k>3$, independent of the choice of $s_0$.
\end{Cor}
\begin{IEEEproof}
Codewords of weight $3$ in $\frac{1}{2}$-rate punctured simplex codes constructed from a $3$-weight parity-check polynomial can only have support given by the indexes of
\begin{enumerate}
    \item three columns of $\mathbf{H}$ of weight $1$, summing up to $0$ modulo $2$;
    \item two columns of $\mathbf{H}$ of weight $1$ and one of weight $2$, summing up to $0$ modulo $2$;
    \item two columns of $\mathbf{H}$ of weight $2$ and one of weight $1$, summing up to $0$ modulo $2$;
    \item three columns of weight $2$ summing up to $0$ modulo $2$.
    \end{enumerate}
    However, since all the columns of $\mathbf{H}$ of weight $1$ have a different support, they cannot sum up to $0$ modulo $2$ and Case 1) is not possible. Case 3) is also impossible, since the sum modulo $2$ of two columns of weight $2$ has either weight $2$ or $4$, and therefore the sum modulo $2$ with a $1$-weight column cannot produce an all-zero vector. In order to study case 4), we notice that the $k$ central columns in Fig. \ref{fig:H3} form a $2k$-length cycle. Since $k>3$ by hypothesis, these columns do not contain $6$-length cycles, and therefore $3$-weight codewords. Therefore, also case 4) cannot occur. In case 2), we have $k$ columns of weight $2$ and for each of them there exists a pair of columns of weight $1$ such that these three columns sum up to $0$ modulo $2$, as shown in the proof of Theorem \ref{the:themindist}. Therefore, $A(3)=k$, independent of the value of $s_0$.
\end{IEEEproof}

The case of a $\frac{1}{2}$-rate punctured simplex code, constructed from a $3$-weight parity-check polynomial for $k=3$ is particular, since the $3$ central columns form a $6$-length cycle, and therefore $A(3)=4$. This $[6,3,3]$ code turns out to be the well-known one-time punctured simplex code, which is self-dual and equivalent to the $1$-time shortened Hamming code, with the same original length $N = 7$, whose overall weight distribution is $A(0) = 1$, $A(3) = 4$, $A(4) = 3$. 

%Now we will discuss briefly higher code rates still with a parity check polynomial $h(x)$ characterized by weight $3$. We will treat the only cases of rate $R = \frac{1}{3}$ and $R = \frac{1}{4}$. When $R = \frac{1}{3}$, the $\mathbf{H}$ matrix, with respect to that depicted in Fig. \ref{fig:H3}, contains $k$ additional rows properly displaced. This implies the introduction of an additional $2k \times k$ central block  formed  by all columns having weight $3$. This leads to the following consequence.

%\begin{Lem}
%In a $\frac{1}{3}$-rate punctured simplex code characterized by a $3$-weight parity check polynomial, among the $3k$ columns of the resulting $\mathbf H$ matrix, $k$ exhibit weight $1$, $k$ weight $2$ and $k$ weight $3$.
%\end{Lem}
%\begin{IEEEproof}
%The proof is a straightforward extension of that of Lemma \ref{lem:lemma12}.
%\end{IEEEproof}

\subsection{Codes characterized by parity-check polynomials with higher weights}\label{sec:hiwe}

Also $5$-weight polynomials, and even $7$-weight polynomials, if $k$ is high enough, can exhibit all different separations (i.e., the associated $\mathbf{p}$ is a Golomb ruler) and hence satisfy the row-column constraint. For example, with $k=16$, the coefficients of the $5$-weight primitive polynomial are $10001000000001011$. It is easy to check that the associated $\mathbf{p}=[0,4,13,15,16]$ is a Golomb ruler. In a similar way, with $k = 121$, the $7$-weight primitive polynomial identified by the following exponents of non-null coefficients $\mathbf{p}=[0,8,25,105,115,116,121]$ is a Golomb ruler. Therefore, owing to Theorem \ref{the:the1}, the row-column constraint is satisfied in both cases. 
Furthermore, qualitatively speaking, the absence of equalities requires the collection of a greater number of columns in the syndrome sum cancellation procedure for finding low-weight codewords, so intrinsically increasing the codeword weight.

\begin{figure}[htbp]
\centering
    \includegraphics[width=0.45\textwidth]{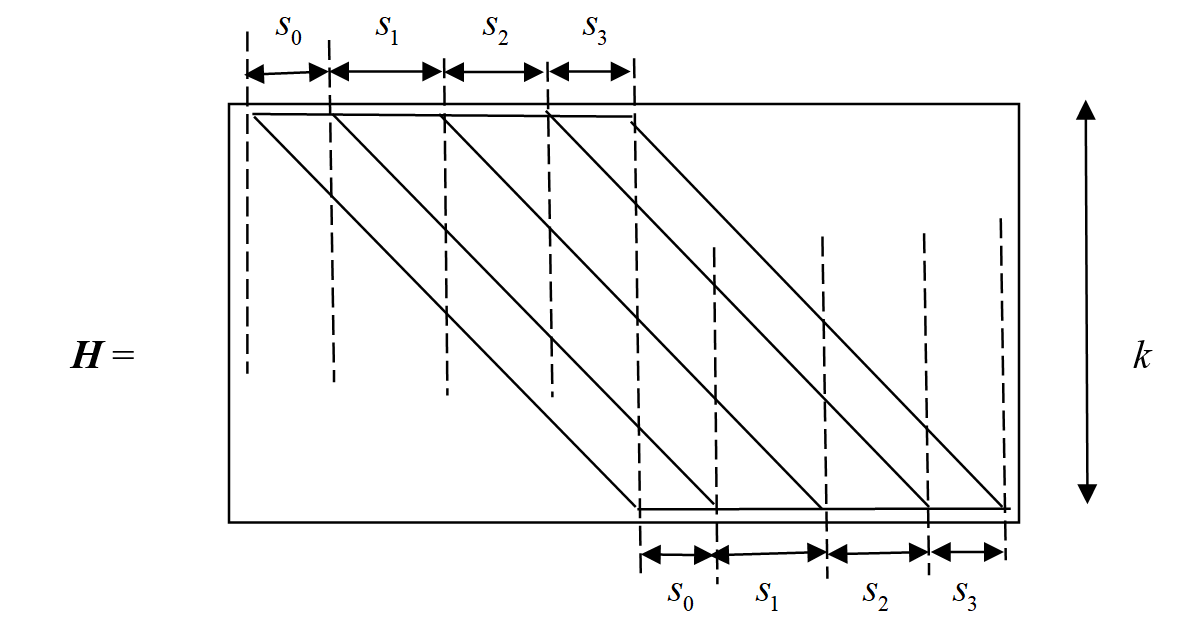}
\caption{Structure of the parity-check matrix for code rate $\frac{1}{2}$ and a $5$-weight parity check polynomial}
\label{fig:H5}
\end{figure}

In Fig. \ref{fig:H5} the structure of the H matrix for code rate $\frac{1}{2}$ and a $5$-weight parity check polynomial is schematically shown. The differences between consecutive elements of $\mathbf{p}$ are named $s_0, s_1, s_2, s_3$ and $s_0+s_1+s_2+s_3=k$. It is possible to draw the following extension of Lemma 2.

\begin{Lem}
In an $\frac{1}{2}$-rate punctured simplex code characterized by a $w$-weight parity-check polynomial, among the $2k$ columns of the resulting $\mathbf{H}$ matrix, $s_i+s_j$ have weight $i+1$ and the same number characterizes those having weight $w-j-1$, where $i\in\{0,\ldots, \frac{w-3}{2}\}$, $j\in \{\frac{w-1}{2},\ldots,w-2\} $ and $i+j=w-2$.
\label{lem:lemma125}
\end{Lem}
\begin{IEEEproof}
Similar to the proof of Lemma \ref{lem:lemma12}.
\end{IEEEproof}

The choice of the three central non-null powers in $h(x)$ influences here the overall performance of the code. In spite of this consideration, the following property is verified about the average column weight $<w_c>$ in the parity-check matrix, which represents an important parameter in order to evaluate decoding complexity.

\begin{The}
In a punctured simplex code having parity-check polynomial weight $w$, length $n=\alpha k$, dimension $k$, code rate $R=\frac{1}{\alpha}=\frac{k}{n}$, the average column weight $<w_c>=w(1-R)$, independent of the vector of differences $\mathbf{s}$.
\end{The}
\begin{IEEEproof}
The thesis follows from the following equality
\[
<w_c>=(\alpha-2)\frac{k}{n}w+\sum_{i=0}^{w-2}s_i\frac{w}{n},
\]
due to the fact that $( \alpha-2)k$ columns have weight $w$ and the remaining ones are as in a $\frac{1}{2}$-rate code. Then, by considering that $k=\sum_{i=0}^{w-2}s_i$, we obtain
\[
<w_c>=(\alpha-2)Rw+Rw,
\]
from which the thesis easily follows.
\end{IEEEproof}

About the expected increase of the minimum distance $d_{\min}$, for a certain choice of $w$ and vector of differences $\mathbf{s}$, as long as the code rate $R$ is reduced with respect to $\frac{1}{2}$, it can be justified by the following qualitative considerations, supported by a polynomial approach \cite{Cancellieri2015}. Owing to the properties of simplex codes, the cofactor $g(x)$ of $h(x)$ with respect to the binomial $x^N+1$ is a long sequence of binary symbols, forming all the possible combinations of $k$ elements except the one formed by $k$ consecutive $0$-symbols. Code rate reduction from $\frac{1}{\alpha}$ to $\frac{1}{\alpha'}$, with $\alpha'=\alpha+1$, implies the addition to all the previous codewords of $k$ consecutive symbols taken from the vector of coefficients of $g(x)$. In such packet of additional $k$ symbols there will be at least one $1$-symbol, so leading to the increment of one unit in the previous minimum distance. Nevertheless, in the next section, some examples of much higher increments will be presented.

\section{Word weight distributions}\label{sec:weidi}

As long as the weight of $h(x)$ increases the performance in terms of minimum distance and word weight distribution $A(w)$ progressively improve. In Table \ref{table:Tabweights}, we show the weight distributions (for low weights) of three codes, all derived from the same $3$-weight $\mathbf{h}=10010001$, which is characterized by $k = 7$, selecting different values of the block length $n$, in order to have code rate $R = \frac{1}{2}, \frac{1}{3}, \frac{1}{4}$. The minimum distance $d_{\min}$ grows from $d_{\min}= 3$ to $d_{\min} = 5$ and finally to $d_{\min} = 9$. Also the asymptotic coding gain, for a soft-decision decoding, that is the parameter $G_\infty = Rd_{\min}$, has been calculated. All the possible polynomials with weight $3$ show the same word weight distribution, up to $A(5)$ at code rate $\frac{1}{2}$. Some small differences appear at lower code rates.

\begin{table*}[h]
\renewcommand{\arraystretch}{1.5}
\caption{Coefficients $A(w)$ of the word weight distributions obtained with the three punctured simplex codes $[14,7,3]$, $[21,7,5]$, $[28,7,9]$ all derived from the same $3$-weight  $\mathbf{h}=10010001$, asymptotic coding gain $G_{\infty}$ and average column weight $<w_c>$.}
%$\frac{g_{\mathrm{free}}-2}{2}m_s+1$
\label{table:Tabweights}
\centering
\begin{tabular}{|c|c|c|c|c|c|c|}
\hline
$n$ & $k$ & $R$ & $d_{\min}$ & $G_{\infty}$ & $A(w)$ & $<w_c>$ \\\hline
$14$ & $7$ & $\frac{1}{2}$  & $3$ & $1.8$ dB & $A(3)=7$, $A(4)=7$, $A(5)=7$, $A(6)=21$ & $1.5$ \\ \hline
$21$ & $7$ & $\frac{1}{3}$  & $5$ & $2.2$ dB & $A(5)=1$, $A(6)=11$, $A(7)=3$, $A(8)=4$ & $2$\\ \hline
$28$ & $7$ & $\frac{1}{4}$  & $9$ & $3.5$ dB & $A(9)=7$, $A(10)=7$, $A(11)=6$, $A(12)=7$ & $2.25$\\ \hline
\end{tabular}
\end{table*}

With $h(x)$ having weight $5$, the particular choice of $h(x)$ can induce remarkable behavior differences. In Table \ref{table:Tabweights2} we have considered a $5$-weight primitive polynomial characterized by $k = 16$. Its binary representation is $10001000000001011$ and the associated $\mathbf{p}$ is a Golomb ruler. In this case, we obtain a minimum distance growing from $d_{\min} = 5$ to $d_{\min} = 10$ and finally to $d_{\min} = 16$. Correspondingly, $G_{\infty}$ increases from $4$ dB to $6$ dB.

\begin{table*}[h]
\renewcommand{\arraystretch}{1.5}
\caption{Coefficients $A(w)$ of the word weight distributions obtained with the three punctured simplex codes $[32,16,5]$, $[48,16,10]$, $[64,16,16]$, all derived from the same $5$-weight  $\mathbf{h}=10001000000001011$, asymptotic coding gain $G_{\infty}$ and average column weight $<w_c>$.}
%$\frac{g_{\mathrm{free}}-2}{2}m_s+1$
\label{table:Tabweights2}
\centering
\begin{tabular}{|c|c|c|c|c|c|c|}
\hline
$n$ & $k$ & $R$ & $d_{\min}$ & $G_{\infty}$ & $A(w)$ & $<w_c>$ \\\hline
$32$ & $16$ & $\frac{1}{2}$  & $5$ & $4$ dB & $A(5)=2$, $A(6)=22$, $A(7)=67$ & $2.5$ \\ \hline
$48$ & $16$ & $\frac{1}{3}$  & $10$ & $5.2$ dB & $A(10)=4$, $A(11)=12$, $A(12)=3$& $3.33$\\ \hline
$64$ & $16$ & $\frac{1}{4}$  & $16$ & $6$ dB & $A(16)=3$, $A(17)=11$, $A(18)=17$ & $3.75$\\ \hline
\end{tabular}
\end{table*}

Considering that we are dealing with very small block lengths, the results obtained appear encouraging. In particular, the code identified by the tern $[64,16,16]$ is characterized by the same parameters as the direct product of a pair of extended Hamming codes $[8,4,4]$, but the coefficient $A(16)$ is here much smaller.

\begin{figure}[htbp]
\centering
    \includegraphics[width=0.45\textwidth]{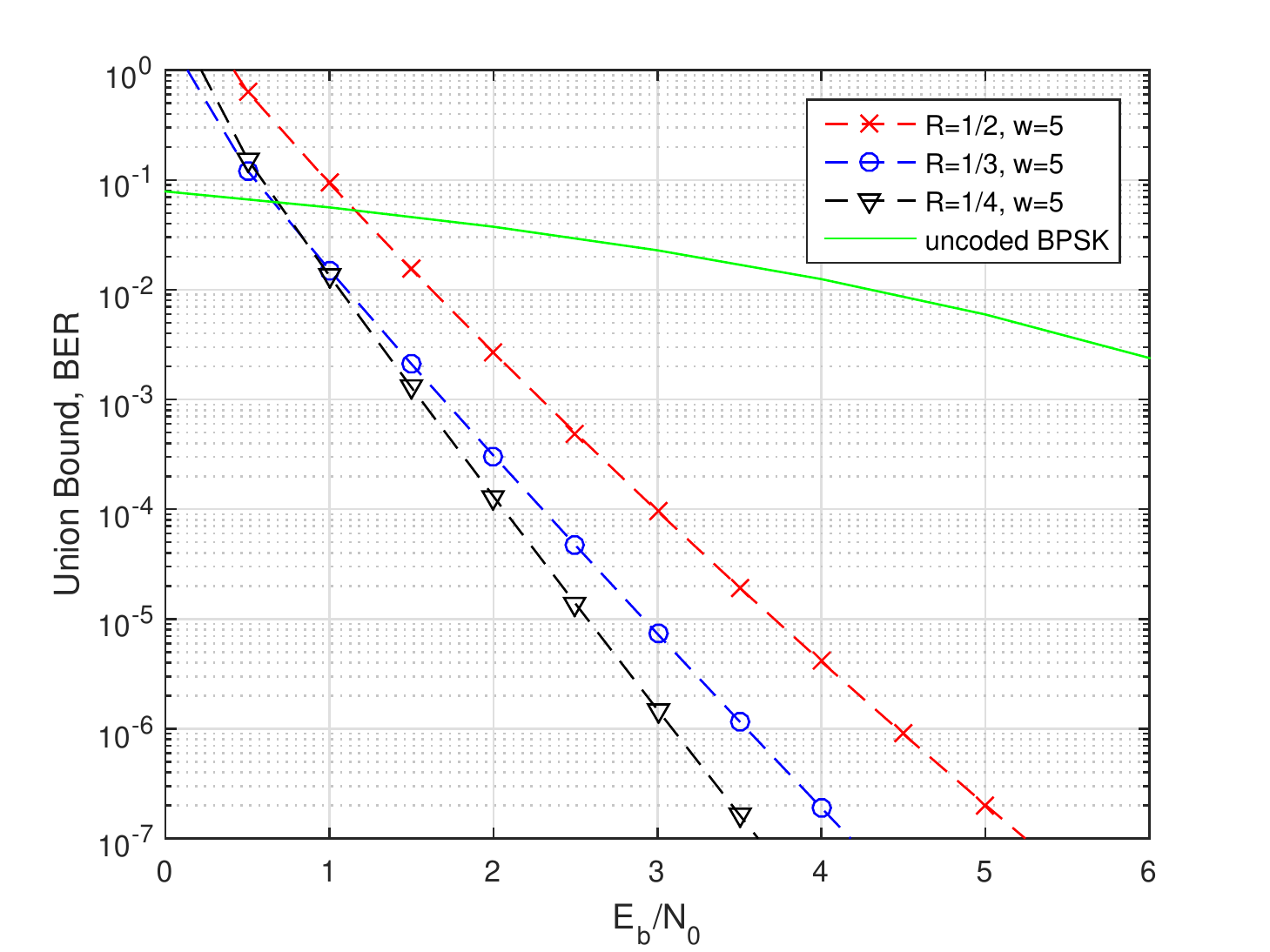}
\caption{Union bound for the codes in Tables  \ref{table:Tabweights2}.}
\label{fig:uniono}
\end{figure}

In both the shown examples, the number of low-weight code words appears very small, yielding a relatively rapid reaching of the asymptotic coding gain $G_{\infty}$. In Fig. \ref{fig:uniono} we show the Truncated Union Bound (TUB) on the Bit Error Rate (BER) 
as a function of the Signal-to-Noise Ratio (SNR), for the codes in Table \ref{table:Tabweights2}, computed as 
\[
\mathrm{BER}_{\mathrm{TUB}}\approx\sum_{w=d_{\min}}^{d^*} \frac{1}{2} \frac{w}{n}A(w)\mathrm{erfc}\left(\sqrt{w\frac{k}{n}\frac{E_b}{N_0}}\right),
\] 
where $d_{\min}\leq d^* \leq n$. Clearly, the larger $d^*$, the tighter the TUB to the complete union bound (obtained for $d^*=n$). The considered values of $d^*$ are $7$, $12$ and $18$, respectively.

\section{Numerical simulations}\label{sec:num}

In this section we have simulated the performance of some punctured simplex codes of rate $\frac{1}{2}$, by means of Monte Carlo simulations of Binary Phase Shift Keying (BPSK) transmissions over an Additive White Gaussian Noise (AWGN) channel. We have adopted a decoding algorithm of the belief propagation family, commonly used to decode LDPC codes. Namely, we have considered the sum-product decoding algorithm \cite{Kschischang}, performing $100$ iterations. The complexity of this algorithm grows linearly with the (average) column weight of the input parity-check matrix (see \cite[Section II]{Battaglioni2018p}).

The first considered code, $C_1$, is defined by a $7$-weight parity-check polynomial, characterized by $k=121$, such that $\mathbf{p}=[0,8,25,105,115,116,121]$, which is a Golomb ruler and, therefore, $C_1$ satisfies the row-column constraint. The second considered code, $C_2$, is defined by a $7$-weight parity-check polynomial, characterized by $k=240$, such that $\mathbf{p}=[0,25,31,138,150,160,240]$, which is a Golomb ruler, too.  

Finally, we have derived a third code, $C_3$, from $C_1$, by applying the so-called circulant expansion technique, which was first proposed in \cite{Tanner2004} (though on different matrices) and further investigated in \cite{RyanBook}. In a nutshell, when using this technique, each $1$-symbol of the starting parity-check matrix is substituted by a circulant permutation matrix of side $p$, and each $0$-symbol is substituted by an all-zero matrix of side $p$. This is known to improve the error rate performance of the starting code (see \cite{Battaglioni2016} for a performance evaluation with increasing values of $p$). In this case, we have used  matrices of side $p=7$. The $i$th $1$-symbol in each row of the parity-check matrix of $C_1$ has been substituted by a circulant permutation matrix, such that the support of its first column is $i$, when $i$ is odd, and $(2i \mod 7)$ when $i$ is even. The resulting code $C_3$ is quasi-cyclic and has dimension $k=121\cdot7=847$. Its parity-check matrix does not contain $4$-length cycles, since the circulant expansion technique preserves the satisfaction of the row-column constraint \cite{MitchellPrelifted}.   

\begin{figure}[htbp]
\centering
    \includegraphics[width=0.45\textwidth]{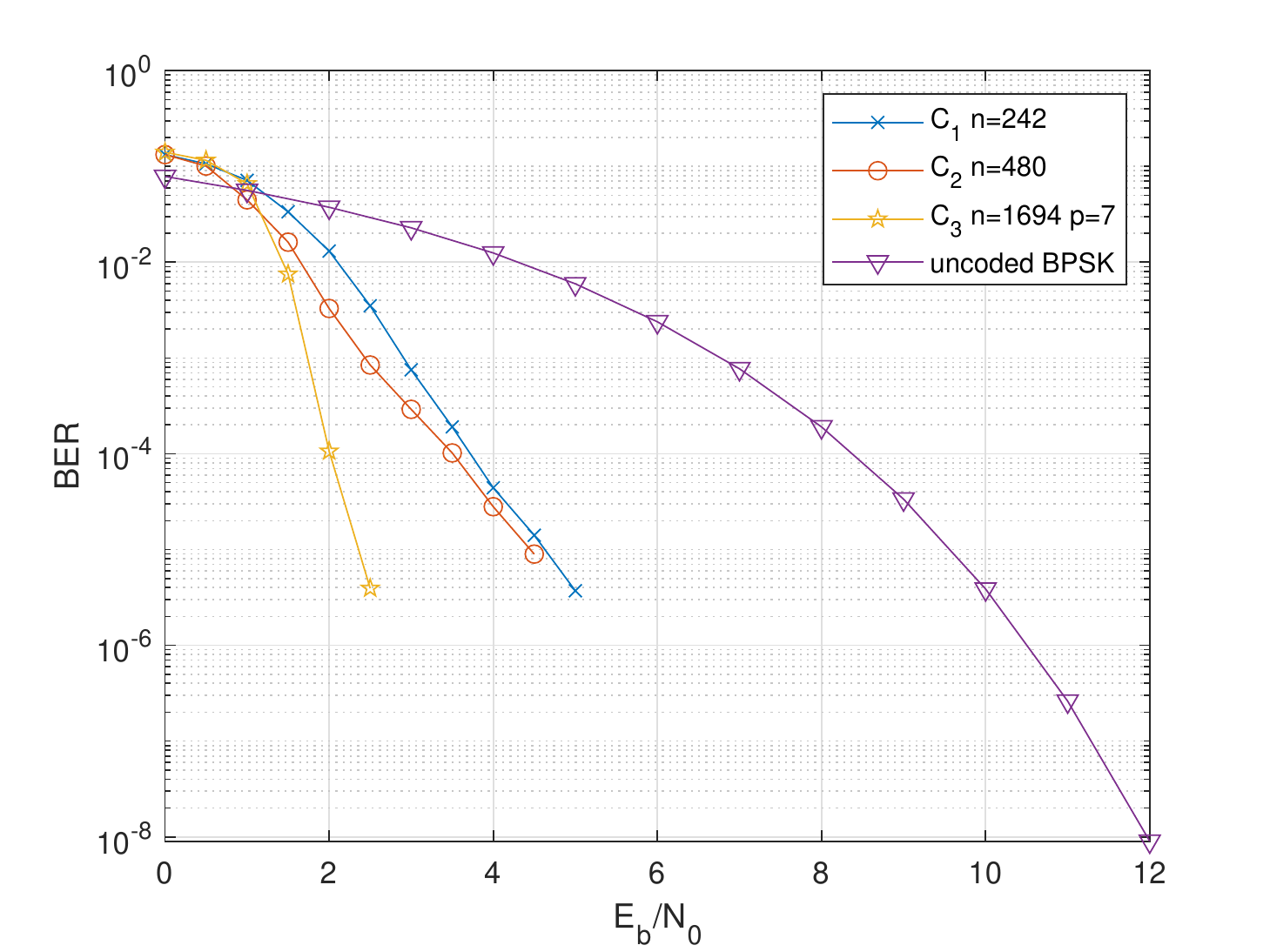}
\caption{Bit error rate performance of codes $C_1$, $C_2$ and $C_3$, as a function of the SNR.}
\label{fig:erp}
\end{figure}

The performance of these codes in terms of BER   is shown in Fig. \ref{fig:erp}. We notice that, with respect to the uncoded case, the proposed codes obtain a relatively large gain. Moreover, Fig. \ref{fig:erp} confirms that punctured simplex codes can be efficiently decoded as LDPC codes, significantly reducing the decoding latency and complexity.

\begin{figure}[htbp]
\centering
    \includegraphics[width=0.45\textwidth]{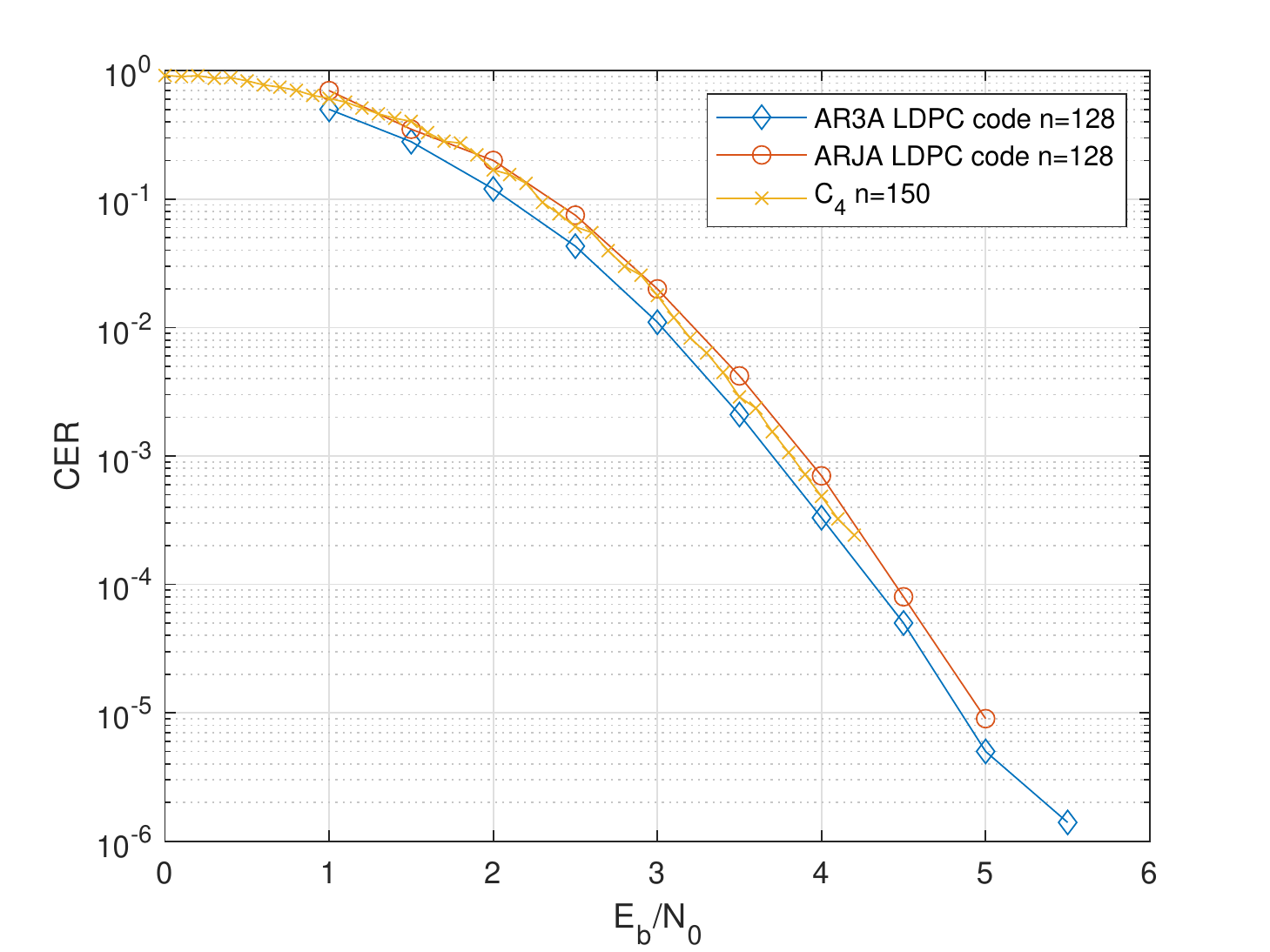}
\caption{Codeword error rate performance of codes $C_4$ and two codes from \cite{Livashort}, as a function of the SNR.}
\label{fig:erpc}
\end{figure}

We have also considered a fourth code, $C_4$, defined by a $7$-weight parity-check polynomial, characterized by $k=75$, such that $\mathbf{p}=[0,2,21,29,60,72,75]$, and compared it with two of the LDPC codes considered in \cite{Livashort}, having $k=64$: an accumulate-repeat-jagged-accumulate (ARJA) LDPC code and an accumulate-repeat-3-accumulate (AR3A) LDPC code. The results are shown in Fig. \ref{fig:erpc}, where we notice that our newly designed code has comparable block length and error rate performance with codes widely employed in standards.

\section{Conclusion}\label{sec:concl}

For any value of the cyclic length $N = 2^k-1$  characterizing a binary simplex code, a family of rate adaptive LDPC codes can be found. This property follows from the existence of a $w$-weight parity-check primitive polynomial. If such a polynomial also corresponds to a Golomb ruler, the parity-check matrix of the code does not contain $4$-length cycles, which are known to degrade the code performance in terms of error rate when belief propagation-based algorithm are used for decoding. If the parity-check polynomial is not associated to a Golomb ruler, the residual 4-length cycles can be eliminated by properly substituting each non-zero symbol in the parity-check matrix with a circulant permutation matrix, leading to a code with larger length but the same code rate. This method might also be used with the purpose of increasing the code minimum distance. So, with code lengths of the order of some thousands, asymptotic coding gains as large as $10$ dB are expected to be reached and even rapidly approached. We leave this analysis for future works.

\bibliographystyle{IEEEtran}
\bibliography{ref}
\end{document}